\documentclass[preprint2]{aastex}
\usepackage{graphicx}

\newcommand{\teff}{\mbox{$T_{\rm eff}$}}
\newcommand{\logg}{\mbox{$\log g$}}
\newcommand{\vsini}{\mbox{$v \sin i$}}
\newcommand{\mictrb}{\mbox{$\xi_{\rm t}$}}
\newcommand{\mactrb}{\mbox{$v_{\rm mac}$}}

\newcommand{\kms}{\mbox{km\,s$^{-1}$}}
\newcommand{\ms}{\mbox{m\,s$^{-1}$}}
\newcommand{\halpha}{\mbox{$H_\alpha$}}

\shorttitle{WASP-41\,b}
\shortauthors{Maxted et~al.}

\begin{document} 
\title {WASP-41\,b: A transiting hot Jupiter planet orbiting a
magnetically-active G8\,V star}

\author{
P.F.L. Maxted\altaffilmark{1}, 
D.R. Anderson\altaffilmark{1}, 
A. Collier Cameron\altaffilmark{2}, 
C. Hellier\altaffilmark{1},
D. Queloz\altaffilmark{3}, 
B. Smalley\altaffilmark{1}, 
R. A. Street\altaffilmark{4}, 
A.H.M.J. Triaud\altaffilmark{3}, 
R.G. West\altaffilmark{5},
M. Gillon\altaffilmark{6}, 
T.A. Lister\altaffilmark{4}, 
F. Pepe\altaffilmark{3}, 
D. Pollacco\altaffilmark{7}, 
D. S\'egransan\altaffilmark{3}, 
A. M. S. Smith\altaffilmark{1}, 
S. Udry\altaffilmark{3}}

\altaffiltext{1}{Astrophysics Group, Keele University, Staffordshire, ST5 
5BG, UK}
\altaffiltext{2}{SUPA, School of Physics and Astronomy, University of St.\
Andrews, North Haugh,  Fife, KY16 9SS, UK}
\altaffiltext{3}{Observatoire astronomique de l'Universit\'e de Gen\`eve
51 ch. des Maillettes, 1290 Sauverny, Switzerland}
\altaffiltext{4}{Las Cumbres Observatory, 6740 Cortona Dr. Suite 102, Santa 
Barbara, CA 93117, USA}
\altaffiltext{5}{Department of Physics and Astronomy, University of 
Leicester, Leicester, LE1 7RH, UK}
\altaffiltext{6}{Institut d'Astrophysique et de G\'eophysique, Universit\'e
de Li\`ege, All\'ee du 6 Ao\^ut, 17, Bat. B5C, Li\`ege 1, Belgium,} 
\altaffiltext{7}{Astrophysics Research Centre, School of Mathematics \& 
Physics, Queen's University, University Road, Belfast, BT7 1NN, UK}

\begin{abstract}
 We report the discovery of a transiting planet with an orbital period of
3.05\,d orbiting the star TYC~7247-587-1. The star, WASP-41, is a moderately
bright G8\,V star (V=11.6) with a metallicity close to solar
([Fe/H]$=-0.08\pm0.09$). The star shows evidence of moderate chromospheric
activity, both from emission in the cores of the Ca\,II H and K lines and
photometric variability with a period of 18.4\,d and an amplitude of about
1\%. We use a new method to show quantitatively that this periodic signal has
a low false alarm probability. The rotation period of the star implies a
gyrochronological age for WASP-41 of 1.8\,Gyr with an error of about 15\%. We
have used a combined analysis of the available photometric and spectroscopic
data to derive the mass and radius of the planet ($0.92\pm0.06M_{\rm Jup}$,
$1.20\pm0.06R_{\rm Jup}$). Further observations of WASP-41 can be used to
explore the connections between the properties of hot Jupiter planets and the
level of chromospheric activity in their host stars.  

\end{abstract}


\section{Introduction}
 §There is continued
interest in finding bright stars that host transiting exoplanets because they can be accurately
characterized and studied in some detail, e.g., the mass and radius of the
planet can be accurately measured. This gives us the opportunity to explore
the relationships between the properties of the planet and its host star,
e.g., the orbital eccentricity, the composition and spectral type of the star,
the age of the system, etc. Given the wide variety of transiting planets being
discovered and the large number of parameters that characterise them,
statistical studies will require a large sample of systems to identify and
quantify the relationships between these parameters. These relationships can
be used to test models of the formation, structure and evolution of short
period exoplanets. 

 Here we report the discovery by the WASP survey  of
a planetary mass companion to the star TYC~7247-587-1. We find that the star
is a G8\,V star showing moderate chromospheric activity. The planet,
WASP-41\,b, is a typical hot Jupiter planet with an orbital period of 3.05\,d.

\begin{figure} 
\plotone{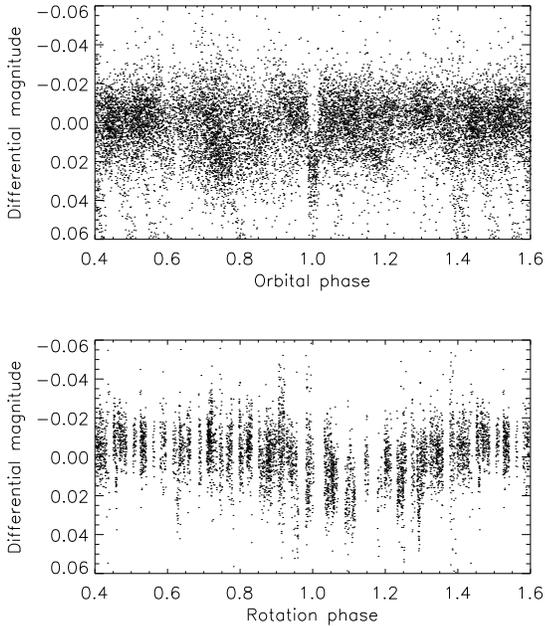}
\caption{WASP-South photometry of WASP-41. Raw data are plotted using small
points, phase-binned data are plotted using filled circles. Upper panel: all
data plotted as a function of the orbital phase with  period $P$~=~3.0524\,d.
Lower panel: data from 2008 plotted as a function  of the rotation phase with
period  $P_{\rm rot}=18.4$\,d. 
 \label{wasplc} }
\end{figure} 
\begin{figure} 
\plotone{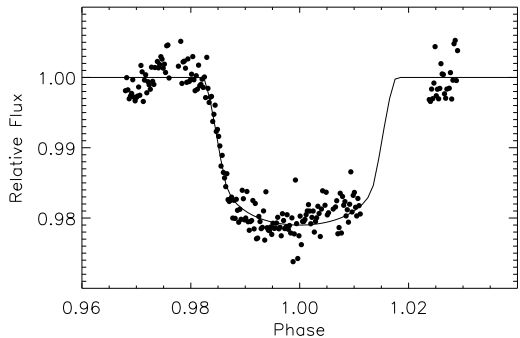} 
\caption{Faulkes Telescope South z-band photometry of WASP-41 (points) with 
 the model fit described in Section~\ref{paramsec} (solid line). 
\label{ftslc} }
\end{figure} 

\begin{figure} 
\includegraphics[width=0.45\textwidth]{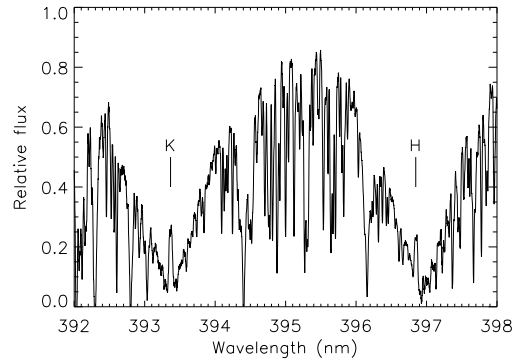} 
\caption{Section of the co-added {\sc coralie} spectra of WASP-41 showing
emission in the core of the Ca\,II H and K lines. \label{hkfig}}
\end{figure} 
\section{Observations}
The WASP survey is described in \citet{2006PASP..118.1407P} and
\citet{2008ApJ...675L.113W} while a discussion of our candidate selection
methods can be found in \citet{2007MNRAS.380.1230C},
\citet{2008MNRAS.385.1576P}, and references therein. 

 The star TYC~7247-587-1 (WASP-41,
 1SWASP J124228.50$-$303823.5) was observed 6767 times by one camera on the
WASP-South instrument from 2007 January 20 to 2007 June 22. A further 5637
observations were obtained with the same camera from 2008 January 17 to 2008
May 28. 

 The WASP-South lightcurves of WASP-41 show transit-like features with a depth
of approximately 0.02 magnitudes recurring with a 3.05-d period
(Fig.~\ref{wasplc}).  These were independently detected in the WASP-South
photometry from the two seasons using the de-trending and transit detection
methods described in \cite{2007MNRAS.380.1230C}, which was taken as  good
evidence that the periodic transit signal was real. The spectral type of the
star was estimated to be approximately G8 based on the catalogue photometry
available for this star at the time. The duration and depth of the transit is
consistent with the hypothesis that it is due to the transit of a planet-like
companion to a main-sequence G8 star and the WASP-South lightcurves show no
indication of any ellipsoidal variation due to the distortion of the star by a
massive companion.

 We obtained 22 radial velocity measurements during the interval 2010 January
3 to 2010 August 5 with the fibre-feb {\sc coralie} spectrograph on the Euler
1.2-m telescope located at La Silla, Chile. The spectra were obtained with an
exposure time of 30 minutes and have  a typical signal-to-noise ratio of
25\,--\,30. Accurate wavelength calibration is ensured by the simultaneous
observation of a thorium-argon arc in a second fibre feed to the spectrograph.
Details of the instrument and data reduction can be found in
\cite{2000A&A...354...99Q} and references therein. The RV measurements were
performed using cross-correlation against a numerical mask generated from a
G2-type star are given in Table~\ref{rv-data}, where we also provide the
bisector span, BS, which measures the asymmetry of the cross-correlation
function \citep{2001A&A...379..279Q}. The standard error of the the bisector
span measurements is $2\sigma_{\rm RV}$. 

 We also obtained photometry of  TYC~7247-587-1 and other nearby stars on 2010
June 23 using the LCOGT 2.0-m Faulkes Telescope South (FTS) at Siding Spring
Observatory. The Merope camera we used has an image scale of 0.279
arcseconds/pixel when used in the 2x2 binning mode we employed. We used a
Pan-STARRS\footnote{\url{http://pan-starrs.ifa.hawaii.edu/public/design-features/cameras.html}}
z-band filter to obtain 210 images covering one transit. These images were
processed in the standard way with IRAF\footnote{IRAF is distributed by the
National Optical Astronomy Observatory, which is operated by the Association
of Universities for Research in Astronomy (AURA) under cooperative agreement
with the National Science Foundation.} using a stacked bias image, dark frame,
and sky flat. The
DAOPHOT photometry package \citep{1987PASP...99..191S} was used to perform
object detection and aperture photometry for WASP-41 and several comparison
stars in the 5'\,$\times$\,5' field of view of the instrument.  Observations
were interrupted by poor weather so there are no observations during the egress phase of
the transit. These data are sufficient to confirm that the transit-like
features seen in the WASP-South data are due to the star TYC~7247-587-1 and to
provide better measurements of the depth of the transit than is possible from
the WASP-South data (Fig.~\ref{ftslc}). 

 All photometric data presented in this paper are available from the NStED
database.\footnote{\url{http://nsted.ipac.caltech.edu}}
\begin{figure*} 
\begin{center}
\includegraphics[width=0.9\textwidth]{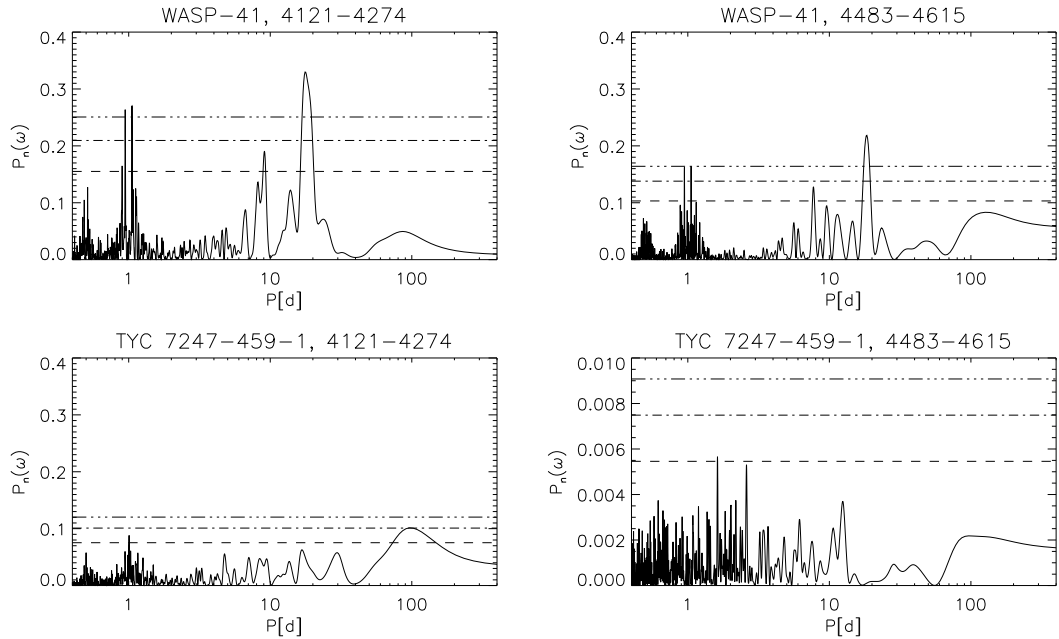}
\end{center}
\caption{Periodogram of the WASP data from two seasons for WASP-41 and
TYC~7247-459-1.   The star name and date range (JD$-$245000) are given in the
title of each panel. Horizontal lines indicate false alarm probability levels
FAP=0.1,0.01,0.001. Note the change of scale in the lower right panel.
 \label{swlomb} }
\end{figure*}

\begin{figure} 
\plotone{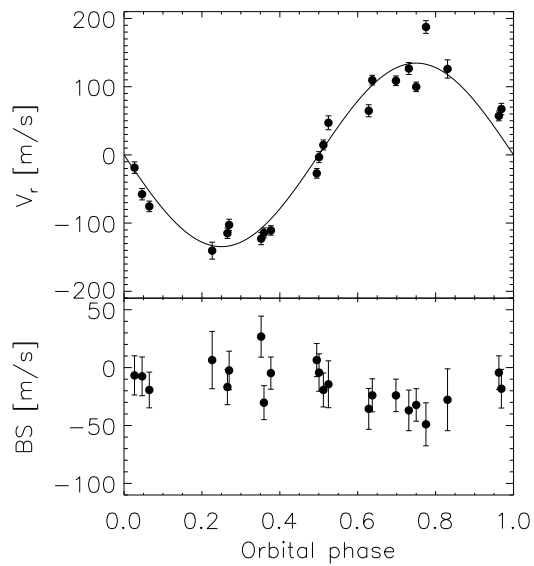} 
\caption{Radial velocity and bisector span measurements for WASP-41. Upper
panel: Radial velocity relative to the centre-of-mass velocity together with
the best-fit circular orbit. Lower panel: bisector span measurements. 
\label{rvphase} }
\end{figure}


\section{WASP-41 Stellar Parameters}

The 17 individual {\sc coralie} spectra of WASP-41 available up to May 2010
were co-added to produce a single spectrum with a signal-to-noise ratio
$\approx$70:1. The standard pipeline reduction products were used in the
analysis.

The analysis was performed using the methods given in
\citet{2009A&A...496..259G}. The \halpha\ line was used to determine the
effective temperature (\teff), while the Na {\sc i} D and Mg {\sc i} b lines
were used as surface gravity (\logg) diagnostics. The parameters obtained from
the analysis are listed in Table~\ref{wasp41-params}. The elemental abundances
were determined from equivalent width measurements of several clean and
unblended lines. A value for micro-turbulence (\mictrb) was determined from
Fe~{\sc i} using the method of \cite{1984A&A...134..189M}. The quoted error
estimates include that given by the uncertainties in \teff, \logg\ and \mictrb,
as well as the scatter due to measurement and atomic data uncertainties.

The projected stellar rotation velocity (\vsini) was determined by fitting the
profiles of several unblended Fe~{\sc i} lines. A value for macro-turbulence
(\mactrb) of 2.3 $\pm$ 0.3 \kms\ was assumed, based on the tabulation by
\cite{2008oasp.book.....G}, and an instrumental FWHM of 0.11 $\pm$ 0.01 \AA,
determined from the telluric lines around 6300\AA. A best fitting value of
\vsini\ = 1.6 $\pm$ 1.1~\kms\ was obtained.

\begin{table} 
\caption{Radial velocity measurements for WASP-41} 
\label{rv-data} 
\begin{tabular}{lrrr} 
\tableline 
\noalign{\smallskip}
BJD&\multicolumn{1}{l}{RV} & 
\multicolumn{1}{l}{$\sigma_{\rm RV}$} & 
\multicolumn{1}{l}{BS}\\ 
$-$2\,450\,000  & (km s$^{-1}$) & (km s$^{-1}$) & (km s$^{-1}$)\\ 
\noalign{\smallskip}
\tableline
\noalign{\smallskip}
 5200.8101 &3.1691 &0.0076  &$-$0.0168 \\
 5290.8090 &3.3837 &0.0070  &$-$0.0323 \\
 5293.8035 &3.4106 &0.0087  &$-$0.0369 \\
 5294.8225 &3.2084 &0.0077  &$-$0.0193  \\
 5295.7751 &3.1732 &0.0069  &$-$0.0048  \\
 5296.7550 &3.3926 &0.0070  &$-$0.0240  \\
 5297.8185 &3.2263 &0.0083  &$-$0.0076  \\
 5298.7732 &3.1696 &0.0073  &$-$0.0303  \\
 5300.8114 &3.2652 &0.0084  &$-$0.0067  \\
 5301.8036 &3.1611 &0.0088  &$+$0.0268  \\
 5305.7269 &3.3935 &0.0071  &$-$0.0239  \\
 5306.7188 &3.3411 &0.0072  &$-$0.0043  \\
 5311.8040 &3.3486 &0.0088  &$-$0.0356  \\
 5317.5927 &3.3310 &0.0101  &$-$0.0144  \\
 5323.6579 &3.2985 &0.0073  &$-$0.0193  \\
 5326.6585 &3.2569 &0.0071  &$+$0.0065  \\
 5327.6832 &3.4099 &0.0133  &$-$0.0278  \\
 5362.4677 &3.1436 &0.0123  &$+$0.0065  \\
 5375.5148 &3.2806 &0.0080  &$-$0.0043  \\
 5388.5621 &3.4714 &0.0092  &$-$0.0490  \\
 5410.5201 &3.3512 &0.0083  &$-$0.0182  \\
 5414.4912 &3.1813 &0.0083  &$-$0.0024  \\
\noalign{\smallskip}
\tableline 
\end{tabular} 
\end{table}

\begin{table}[h]
\caption{Stellar parameters of WASP-41 from spectroscopic analysis.}
\begin{tabular}{lr} \hline
Parameter  & Value \\ \hline
\teff ~[K]  & 5450 $\pm$ 150  \\
\logg      & 4.4 $\pm$ 0.2 \\
\mictrb ~ [\kms]   & 1.0 $\pm$ 0.2 \\
\vsini  ~ [\kms]   & 1.6 $\pm$ 1.1 \\
{[Fe/H]}   &$-$0.08 $\pm$ 0.09 \\
{[Na/H]}   &   0.07 $\pm$ 0.09 \\
{[Mg/H]}   &   0.07 $\pm$ 0.16 \\
{[Al/H]}   &$-$0.01 $\pm$ 0.08 \\
{[Si/H]}   &   0.05 $\pm$ 0.06 \\
{[Ca/H]}   &   0.08 $\pm$ 0.15 \\
{[Sc/H]}   &   0.01 $\pm$ 0.10 \\
{[Ti/H]}   &   0.00 $\pm$ 0.10 \\
{[V/H]}    &   0.06 $\pm$ 0.17 \\
{[Cr/H]}   &   0.00 $\pm$ 0.05 \\
{[Mn/H]}   &   0.00 $\pm$ 0.13 \\
{[Co/H]}   &$-$0.01 $\pm$ 0.07 \\
{[Ni/H]}   &$-$0.04 $\pm$ 0.06 \\
$\log$ A(Li)  &   $<$ 0.5 \\
Mass  ~ [$M_{\sun}$]    &   0.95 $\pm$ 0.09  \\
Radius~ [$R_{\sun}$]    &   1.01 $\pm$ 0.26  \\
Sp. Type   &   G8 V\\
Distance [pc]  &   180 $\pm$ 60 \\ \hline
\\
\end{tabular}
\label{wasp41-params}
\newline {\bf Note:} Mass and Radius estimate using the
\cite{2010A&ARv..18...67T} calibration. Spectral Type estimated from \teff\
using the table in \cite{2008oasp.book.....G}.
\end{table}


 We observe emission lines in the cores of the Ca\,II H and K lines in our
spectrum of WASP-41 (Fig.~\ref{hkfig}). We estimate a spectral index $\log
R'_{HK} \approx -$4.67 from this spectrum \citep{1984ApJ...279..763N}, but
note that a transformation for measurements made with the {\sc coralie}
spectrograph to a standard system does not yet exist.

\section{Rotation period}
 The presence of Ca\,II H and K emission lines in the spectrum of WASP-41
suggests the possibility of variations in the lightcurve due to
star spots with the same period as the rotation period of the star. The
projected equatorial  rotation velocity of WASP-41 combined with the estimated
radius given in Table~\ref{wasp41-params} imply a rotation period of
$32\pm23$\,d. We therefore searched for photometric variations in our WASP
photometry with a similar period.

 Our WASP photometry has a large number of observations (several thousand)
with standard errors that vary from a few milli-magnitudes for the best data
up to a magnitude for data obtained on cloudy nights. WASP data obtained on cloudy
nights is sometimes affected by large systematic errors, so we remove data
with standard errors larger than 5 times the median value prior to further
analysis. The remaining data still have a range of standard errors to it is
advantageous to use a period searching algorithm that can account for this. 
 The generalized Lomb-Scargle periodogram defined by
\citet{2009A&A...496..577Z} is a suitable method. However, there is little to
be gained from including a ``floating mean'' in the case of the WASP data, so we
used a slightly different definition of the periodogram equivalent to a
least-squares fit of the sinusoidal function $y_i = a\sin(\omega t_i) +
b\cos(\omega t_i)$ to magnitudes $m_i=1,2,\dots,N$ with with standard errors
$\sigma_i$ obtained at times $t_i$ for a given angular frequency $\omega$.
Note that we implicitly assume that the magnitudes $m_i$ have a weighted mean
value of 0.  We used the following definition of the power at angular
frequency $\omega$,
 \[ P_n(\omega) = \frac{\chi^2_0- \chi^2(\omega) }{\chi^2_0}, \]
 where
 \[\chi^2_0 = \sum \frac{m_i^2}{\sigma_i^2}\] 
and
\[\chi^2(\omega) = \sum \frac{(m_i-y_i)^2}{\sigma_i^2}.\] 

We used this definition of the power to search for periodicity in the
WASP-South lightcurves of WASP-41 at 4096 evenly spaced frequency values
from  0 to 2.5 cycles/day.   Variability due to star spots is not expected to be
coherent on long timescales as a consequence of the finite lifetime of
star-spots and differential rotation in the photosphere so we analysed the two
seasons of data for WASP-41 separately. The signal of the transits in the data
was removed using a model similar to the one described below, although in
practice this has a negligible effect on the periodogram. We also analysed 3
nearby stars of similar magnitude and colour observed simultaneously with the
same camera. The results are shown in Table~\ref{periodtable}. The
periodograms of WASP-41 and one of the nearby stars are shown in
Fig.~\ref{swlomb}.

 We used the method of \citet{1989ApJ...338..277P} to allow for fast and
accurate computation of the periodogram. This enabled us to investigate the
significance levels of any peaks in our periodograms using a boot-strap
Monte-Carlo approach. We generated synthetic data sets using a method similar
to that devised by \citet{2009MNRAS.400..451C}, in which data are  shuffled
according to the date on which the data were obtained. The shuffling procedure
shifts each night's observations in their entirety to a new date. This
procedure effectively destroys coherent signals with periods longer than 1 day,
but retains the global form of the window function and the effects of
correlated noise. In additional to shuffling the date of the observations, we
also change the sign of all the data from a given night for a random selection
of half of the dates of observation. 

 \citet{2009MNRAS.400..451C} use a single simulated data set to estimate the
parameter $N_{\rm eff}$ in the following model for the distribution of power
in the periodogram in the absence of any periodic signal; \[{\rm Prob}(P_n >
P_n^{\prime}) = \left(1 - P_n^{\prime}\right)^{(N_{\rm eff}-3)/2}.\] They then
use this model to estimate the false alarm probability, FAP, for the highest
peak in the periodogram of the actual data from \[ {\rm FAP} = 1 -
\left[1-{\rm Prob}(P_n > P_n^{\prime})\right]^M, \] where
$M=(t_N-t_1)(f_{\rm max}-f_{\rm min})$ is  an estimate of the number of
independent frequencies in the periodogram calculated over the range of
frequencies $f_{\rm min}$ to $f_{\rm max}$.\footnote{Note that there is
an error in equation (2) of \citeauthor{2009MNRAS.400..451C}, the equation for
FAP given here is the one actually used in that paper.} We found from
simulations of a large number of data sets using both the ``night-shuffling''
method and Gaussian random noise that this method gives inaccurate estimates
of FAP, particularly when this value is small. The principle reason for this
is that the value of N$_{\rm eff}$ estimated from a single periodogram has a
rather large uncertainty, typically around 20\% for the data sets we
investigated. We found that the following empirical model gives a good
representation of the distribution of the peak power $P_{\rm n,best}$ in the
periodograms we generated from simulated data \[ \begin{array}{ll} \log_{10}
[{\rm Prob}(P_{\rm n,best} > P_{\rm n,best}^{\prime} )] = & c_0 + c_1 P_{\rm
n,best}^{\prime} \\ &+ c_2\left(P_{\rm n,best}^{\prime}\right)^2 \\
\end{array} \] For the analysis of a given set of data we  use this model to
fit the distribution of peak power values in 1024 simulated data sets. We then
use the values of $c_0$, $c_1$ and $c_2$ to calculate the power corresponding
to FAP=0.1, 0.01 and 0.001 shown in Fig.~\ref{swlomb}  and to estimate the FAP
value for the highest peak in the actual periodogram given in
Table~\ref{periodtable}.

 There is a clear peak in the periodogram for WASP-41 near P=18\,d, although
there appears to be a difference in the period derived from the two
seasons of data (17.6\,d and 18.4\,d). We inspected the periodograms from the first season of data
for WASP-41 and the three nearby stars and found that they all show power near
1\,d, 30\,d and various combinations of these frequencies and their harmonics,
presumably as a result of systematic errors in the photometry related to the
diurnal and lunar cycles.  In contrast, there is very little spurious
power in the periodograms for the second season of data. For this reason we
identify the period of 18.41$\pm 0.05$\,d derived from the second season of
data as the correct rotation period for WASP-41. We estimated the standard
error of the period measurement by analysing 1024 data sets with the same
number of points as the original sample, randomly re-sampled with reselection
from the original data.

\begin{table*} 
\caption{Period analysis of WASP-42 and nearby stars of similar magnitude and
colour. }
\label{periodtable} 
\begin{tabular}{lrrrrrrrr} 
\tableline 
\noalign{\smallskip}
Star&\multicolumn{1}{l}{V} & 
\multicolumn{1}{l}{B$-$V} &
\multicolumn{1}{l}{Year}& 
\multicolumn{1}{l}{N} &
\multicolumn{1}{l}{RMS} &
\multicolumn{1}{l}{P$_{\rm best}/d$}&
\multicolumn{1}{l}{Power}&
\multicolumn{1}{l}{FAP}\\
\noalign{\smallskip}
\tableline
\noalign{\smallskip}
WASP-41 &11.6&0.7&2007&5673&0.041&17.62&0.358&$2\times10^{-5}$\\
        &    &   &2008&4966&0.019&18.41&0.240&$6\times10^{-7}$ \\ 
TYC~7247-1008-1&11.8&0.3&2007&5414&0.026& 1.00&0.167&0.07 \\
               &    &   &2008&4834&0.012& 0.98&0.023&0.65\\   
TYC~7247-459-1 &11.5&0.5&2007&5389&0.025&96.38&0.101&0.07 \\
               &    &   &2008&5003&0.012& 1.62&0.006&0.45  \\   
TYC~7247-683-1 &11.2&0.5&2007&5325&0.024&86.23&0.089&0.06 \\
               &    &   &2008&4993&0.012& 0.46&0.005&1.00 \\   
\noalign{\smallskip}
\tableline 
\end{tabular} 
\end{table*}

\section{Planetary parameters\label{paramsec}}

 Our radial velocity (RV) and bisector span (BS) measurements for WASP-41 are
shown in  Fig.~\ref{rvphase} as a function of the photometric transit phase.
There is a weak but statistically significant anti-correlation between these
RV and BS values. In principle, such an anti-correlation can be used to
identify spurious RV signals caused by stellar activity. For
example, \citep{2001A&A...379..279Q} showed that the apparent RV
signal in HD166435 is caused by stellar activity by identifying 3
characteristics of the signal: i. the RV signal was not coherent
over timescales of more than 30 days; ii. the amplitudes of the BS and RV
variation were similar; iii. the BS variations were correlated with the
photometric variations of the star. In the case of WASP-41 there is no doubt
that the RV signal is due to the presence of a planetary mass companion to
WASP-41 and not due to stellar activity because: i. the RV signal
is coherent over a timescale of more than 200 days; ii. the amplitude of the
BS variation is an order of magnitude smaller than the RV variation; iii. the
BS values are also correlated with the phase calculated for a rotation period
of 18.4d established from the photometric variations of WASP-41. As an
additional test of the reality of the planetary RV signal we also re-measured
the RV values using a K5-type mask instead of a G2-type mask. Looking for a
difference between the RV values derived with the different masks is an
effective method for detecting spurious RV signals caused by star spots or
faint eclipsing binary stars blended with the target star. In the case of
WASP-41 there is no significant difference in the RV values measured with the
two masks so the best explanation for the observed RV and BS signals is that
WASP-41 has a planetary mass companion with an orbital period of 3.05\,d.

 The {\sc coralie} radial velocity measurements were combined with the
WASP-South and FTS photometry in a simultaneous Markov-chain Monte-Carlo
(MCMC) analysis to find the parameters of the WASP-41 system. We removed the
photometric variation due to stellar rotation from the WASP-South photometry by
subtracting a first-order harmonic series, fit by least-squares to the
differential magnitudes  from each season independently. The shape of the
transit is not well defined in the WASP-South or FTS photometry, so we have
imposed an assumed main-sequence mass-radius relation as an additional
constraint in our analysis of the data. The  stellar mass is determined from
the parameters \teff, \logg\ and [Fe/H] using the procedure described by
\cite{2010A&A...516A..33E}, based on the compilation of eclipsing binary data
by \cite{2010A&ARv..18...67T}. The code uses \teff\ and [Fe/H] as MCMC jump
variables, constrained by Bayesian priors based on the
spectroscopically-determined values given in Table~\ref{wasp41-params}.
Limb-darkening coefficients are taken from \cite{2000A&A...363.1081C}. 

 The parameters derived from our MCMC analysis assuming a circular orbit are
listed in Table~\ref{sys-params}. We found that the contribution to the value
of $\chi^2$ for the fit from the RV data was much higher than
expected given the standard errors of these measurements. This is a result of
additional noise in the RV data due to stellar activity
(``jitter''). The free parameters in the fit to the RV are $T_0$,
$P$, $\gamma$ and $K$, but the values of  $T_0$ and $P$ are determined almost
entirely by the photometric data, so the number of degrees of freedom in the
fit to the RV data is approximately 20. To achieve a contribution
to the value of $\chi^2$  from our 22 RVs of approximately 20 we found that
we needed to add  21\,\ms\ in quadrature to the standard errors for the RV
data. We also performed an MCMC analysis of the data including the
parameters $\sqrt{e}\cos\omega$ and $\sqrt{e}\sin\omega$ as free parameters.
These parameters are used to describe an eccentric orbit solution because they
are not strongly correlated and are equivalent to a assuming uniform prior
distribution for the value of the eccentricity, $e$. With this eccentric-orbit
solution and the same value of the jitter as the circular orbit solution we
find $e=0.10\pm0.06$. As this is consistent with a circular orbit we adopt the
parameters from the circular orbit solution. As we do not know {\it a priori}
that the orbit is circular, we take the standard errors on the parameters from
the non-circular orbit solution.

 We considered the contribution of correlated errors (``red noise'') to the
standard errors quoted in Table~\ref{sys-params}. While the individual
transits in the WASP data are affected by red noise, the analysis of the
combined lightcurve covering many individual transits will not be strongly
affected by red noise because there will be no correlation between the
systematic noise from different nights. The FTS lightcurve is affected by red
noise so we have investigated the effect of this using the ``prayer bead''
method. A separate MCMC analysis was performed in which synthetic FTS
lightcurves were created from the model fit to the lightcurve and the
residuals from this model after cyclic permutation at each step in the MCMC
chain. We find that this does not significantly increase the error estimates
for any of the parameters.

 Another issue to consider for the FTS lightcurve is the effect of
starspots on the lightcurve, particularly given the sparse phase coverage of
this lightcurve. Star-spots can affect the parameters derived from this
lightcurve in two ways. Firstly, the normalization of the lightcurve can be
affected by the overall change in brightness of the system due to the 18.4\,d  
rotational modulation of the lightcurve. Secondly, small star
spots covered  by the planet during the transit by distort the shape of the
lightcurve in a way that may be hard to spot by-eye, but that introduce a
systematic error in the parameters derived \citep{2008ApJ...682..593M}.
We have identified parameters in our solution that are determined primarily by
the FTS lightcurve by performing an MCMC solution using the WASP and {\sc
coralie} data only. We find that the depth of transit and the planet radius
are the only parameters for which the standard error estimates increase
significantly in the solution excluding the FTS data. In both cases, the
values from the two solutions agree, but the solution excluding the FTS data
has standard error estimates approximately twice as large as the solution
including the FTS data. We conclude that these two parameters may be subject
to systematic errors comparable to the random errors quoted in
Table~\ref{sys-params}.  

 We have compared the values of the stellar effective temperature, $T_{\rm
eff}$,  and the stellar density, $\rho_{\star}$ to the stellar models of
\cite{2000A&AS..141..371G}. We use the parameters \teff\ and $\rho_{\star}$
because they are independently determined directly from the observations. We
find that the mass inferred from the models ($0.9\pm0.1M_\sun$) is consistent
with the mass derived in our MCMC analysis and that  the uncertainties on the
values of \teff\ and $\rho_{\star}$  are too large for the models to provide any
useful constraint on age of the star. The surface gravity derived from our
MCMC solution is consistent with the \logg\ value from the analysis of the
spectrum, but the large uncertainty on the latter value means that this is a
rather weak constraint.

\section{Discussion}
 WASP-41\,b joins a growing number of planets discovered with
masses $\approx 0.9M_{\rm Jup}$, radii $\approx 1.2R_{\rm Jup}$ orbiting
solar-like stars with periods of about 3 days, e.g., HAT-P-13\,b, XO-1\,b,
WASP-28\,b, CoRoT-12\,b, WASP-26\,b, HAT-P-5\,b, HAT-P-6\,b, etc.\footnote{\it
http://exoplanet.eu} The WASP-41 planetary system is also similar to the 
TrES-1 system, particularly in regard to the activity level of the host star. 

 \cite{2010ApJ...720.1569K} suggest that there is a clear connection between
the properties of the emission spectra  from hot Jupiters and the activity
levels of their host stars. In general, the emission spectrum of hot Jupiters
measured from the eclipse depths at infrared wavelengths can only be matched
by models that include a high-altitude temperature inversion. Four exceptions
to this general rule are the planets orbiting the stars HD\,189733, TrES-3,
TrES-1 and WASP-4. \citeauthor{2010ApJ...720.1569K} found that all four of
these stars show moderate levels of chromospheric activity, higher than for
all the other stars for which they were able to measure a value $\log
R'_{HK}$. The available evidence supports a hypothesis in which some source of
optical opacity high in the atmosphere of hot Jupiter planets causes a
temperature inversion, but that this opacity source is destroyed by the UV
flux associated with chromospheric activity. The result found by
\citeauthor{2010ApJ...720.1569K} has a high statistical significance but is
based on a sample of only 15 stars. The situation is further complicated by
the correlation between a planet's surface gravity and the level of
chromospheric activity in its host star claimed by \cite{2010ApJ...717L.138H}.
The values of $\log R'_{HK}$ and $\log g_p$ we have measured for WASP-41 are
consistent with the correlation observed by \citeauthor{2010ApJ...717L.138H}.
The connection between chromospheric activity and temperature inversions can
be confirmed in the case of WASP-41 using observations of the secondary
eclipse with the IRAC instrument on the Spitzer Space Telescope at  3.6$\mu$m
and 4.5$\mu$m \citep{2010ApJ...720.1569K}.

 In principle, one can use an age\,--\,activity relation established from the
average behaviour of many solar-type stars to estimate the age of a star based
on the measured $\log R'_{HK} $ value. However, solar-type stars have
activity cycles with periods of about a decade during which the value of $\log
R'_{HK} $ can vary by 10\% or more \citep{1995ApJ...438..269B}. This means
that an age estimate based on a single measurement of $\log R'_{HK} $ has an
unknown systematic error that can be large enough to make the age estimate
effectively meaningless. A more useful age estimate can be made in this case
based on the observed rotation period of the star, $P_{\rm rot}=18.4$\,d. The
calibration of the ``gyrochronological'' age given by
\cite{2007ApJ...669.1167B} implies an age of 1.8\,Gyr for WASP-41 with an
error of about 15\%. Stars with \teff=5300\,--\,5600\,K are seen to be lithium
poor in clusters with ages $>1$\,Gyr such as NGC752, M67 but the majority of
stars in this \teff\ range in clusters with ages of about 600\,Myr such as
Praesepe and the Hyades, NGC6633 and Coma Berenices are lithium rich.  We
conclude that the upper limit to the lithium abundance given in
Table~\ref{wasp41-params} is consistent with any age greater than about
600\,Myr \citep{2005A&A...442..615S}.

\acknowledgments
WASP-South is hosted by the South African Astronomical Observatory and we are
grateful for their ongoing support and assistance. Funding for WASP comes from
consortium universities and from the UK's Science and Technology Facilities
Council. We thank the referee for their careful reading of the manuscript and
for comments that improved the quality of this paper.

\begin{table*} 
\begin{center} 
\caption{System parameters for WASP-41. The planet equilibrium temperature is
calculated assuming a value for the Bond albedo A$=0$. {\bf N.B.} an assumed
main-sequence mass-radius relation is imposed as an additional constraint in
this solution so the mass and radius of the star are not independent
parameters -- see \cite{2010A&A...516A..33E} for details. Parameter values are
taken for the solution assuming $e=0$, standard errors on the parameters are
taken from the solution with $\sqrt{e}\cos(\omega)$ and $\sqrt{e}\sin(\omega)$
as free parameters. } 
\label{sys-params} 
 \begin{tabular}{lcrl}
\noalign{\smallskip}
\tableline\tableline
 Parameter & Symbol & Value & Units \\
 \tableline
 Transit epoch (HJD) & $T_0$ & $  2455343.463 \pm  0.001 $ & days \\ 
 Orbital period & $P$ & $   3.052401  \pm  0.000004 $ & days \\ 
 Planet/star area ratio$^a$ & $(R_p/R_*)^2$ & $  0.0186 \pm  0.0004 $ & \\ 
 Transit duration & $t_T$ & $  0.108 \pm  0.002 $ & days \\ 
\noalign{\smallskip}
 Impact parameter & $b$ & $  0.40 ^{+0.15}_{-0.11}$ & $R_*$ \\ 
\noalign{\smallskip}
 Stellar reflex velocity & $K_1$ & $  0.135\pm 0.008$ & \kms \\ 
\noalign{\smallskip}
 Centre-of-mass velocity & $\gamma$ & $   3.284 \pm 0.009$ &\kms \\ 
\noalign{\smallskip}
 Orbital eccentricity & $e$ & $ 0$ (fixed)  & \\ 
\noalign{\smallskip}
 Orbital inclination & $i$ & $   87.7\pm 0.08$ & degrees \\ 
\noalign{\smallskip}
 Stellar density & $\rho_{\star}$ & $1.27\pm 0.14$ & $\rho_\sun$ \\ 
\noalign{\smallskip}
 Stellar mass & $M_*$ & $  0.93 \pm  0.03$ & $M_\sun$ \\ 
 Stellar radius & $R_*$ & $  0.90 \pm 0.05 $ & $R_\sun$ \\ 
 Orbital semi-major axis & $a$ & $  0.0402 \pm0.0005 $ & $AU$ \\ 
 Planet radius$^a$ & $R_p$ & $   1.20 \pm  0.06$ & $R_J$ \\ 
 Planet mass & $M_p$ & $  0.92\pm 0.06 $ & $M_J$ \\ 
 Planet surface gravity & $\log g_p $ & $ 3.16 \pm 0.04 $ & [cgs] \\
 Planet density & $\rho_p$ & $  0.50\pm0.08 $ & $\rho_J$ \\ 
 Planet temperature & $T_{\rm eq}$ & $   1230 \pm 50$ & K \\ 
 \tableline
\end{tabular} 
\end{center} 
$^a$ These quantities may have a systematic error comparable to the random
error due to the influence of stars spots on the FTS lightcurve.
\end{table*}

\bibliographystyle{pasp}
\bibliography{wasp}

\end{document}